\documentclass{nature}

\usepackage[T1]{fontenc} 
\usepackage{amsmath, amsthm, amssymb, amsfonts, verbatim}
\usepackage{bm}
\usepackage[section]{placeins}
\usepackage{hyperref}

 \allowdisplaybreaks

\newcommand{\rhoHat}{\hat{\varrho}_j}
\newcommand{\rhoHatSub}[1]{\hat{\varrho}_{#1}}

\newcommand{\PP}{\mathbb{P}}

\title{Strong Sure Screening of Ultra-high Dimensional Categorical Data}

 \author{{Randall} {Reese}$^1$, {Xiaotian} {Dai}$^1$, \& {Guifang} {Fu}$^1$}
 
\begin{document}

\maketitle
\begin{affiliations}
 \item Department of Mathematics and Statistics, Utah State University
\end{affiliations}



\begin{abstract}
Feature screening for ultra high dimensional feature spaces plays a critical role in the analysis of data
sets whose predictors exponentially exceed the number of observations. Such data sets are becoming
increasingly prevalent in areas such as bioinformatics, medical imaging, and social network analysis.
Frequently, these data sets have both categorical response and categorical covariates, yet extant feature
screening literature rarely considers such data types. We propose a new screening procedure rooted in
the Cochran-Armitage trend test. Our method is specifically applicable for data where both the
response and predictors are categorical. Under a set of reasonable conditions, we demonstrate that our
screening procedure has the strong sure screening property, which extends the seminal results of Fan
and Lv. A series of four simulations are used to investigate the performance of our method relative to
three other screening methods. We also apply a two-stage iterative approach to a real data example by
first employing our proposed method, and then further screening a subset of selected covariates using lasso, adaptive-lasso and 
elastic net regularization.
\end{abstract}

\section{Introduction}
With the ever increasing prevalence of high and ultra-high dimensional data in fields such as bioinformatics, medical imaging and tomography, 
finance, and sensor systems, there has arisen an accompanying need for methods of analyzing said data. 
Developing methods for the analysis of such data requires methods that are not only statistically sound and accurate, but that moreover are 
computationally tractable. 
\cite{Donoho} provides us with a holistic overview of challenges in high dimensional data analysis. 
\cite{FanLi:BigData2006} and \cite{FanHanLiu:BigData} expand upon the statistical challenges of high dimensional data analysis. 

One fundamental pursuit that has received considerable attention in recent literature is variable selection or feature screening. Based on the 
concept of sparsity, feature screening aims to select a relatively small set of important variables from an overall large feature space. For lexical consistency, given $n$ samples for each of $p$ variables, we will use the term ``high dimensional" to mean $p = \mathcal{O}(n^\xi)$ for some  $\xi>0$, and the term ``ultra-high dimensional" to mean $\log(p) = \mathcal{O}(n^\xi)$ for some  $\xi>0$.

A fundamental challenge of variable selection in high and ultra-high dimensional feature spaces comes from the existence of an immense amount of 
noise features. 
This preponderance of noise can lead to an accumulation of aggregate error rates for certain selection methods. For example, as discussed in 
\cite{FanFanHiDim:2008}, when using a discriminant analysis rule such
as LDA or QDA, the population mean vectors are estimated from
the observed sample. In cases where the dimensionality is high, although individual components of the
population mean vectors can be estimated with sufficient accuracy, the aggregated estimation
error can be very large. This will obviously adversely affect the misclassification rate. 

Cases such as the one discussed above introduce us to the motivation behind dimension reduction techniques like feature screening. A 
multiplicity of methods for variable selection in high dimensional feature spaces have been proposed. Methods such as ridge regression 
[\cite{FrankFriedmanRidge}] and LASSO [\cite{Tibshirani}] were early methods that employed penalized least squares. Similar penalized pseudo-likelihood methods such as the smoothly clipped absolute deviation (SCAD) method [\cite{AAJF2001}] the least angle regression (LARS) algorithm [\cite{efron2004}] and the Dantzig selector [\cite{CandesTao}] soon thereafter followed. 

However, as \cite{FanSamworthWu:2009} point out, the computation inherent in these aforementioned methods impedes our ability to directly apply 
them to ultra-high dimensional feature spaces. 
The simultaneous challenges of computational expediency, statistical accuracy, and algorithmic
stability often make such approaches intractable.

In their pioneering paper, \cite{FanLv:2008} lay the ground work for sure independent screening (SIS) feature screening in ultra-high dimensional feature spaces and established the conceptual underpinnings of much of the literature that would thereafter follow. 
This new era of research sought to overcome the computational limitations of the previous approaches and develop a repertoire of methods viable for the rapidly growing (both in size and totality) ultra-high dimensional data sets requiring analysis.

 Most early approaches stemming from \cite{FanLv:2008} were constructed under assumptions on various forms of linear models between the response 
 and the covariates. In that original paper itself, Fan and Lv assumed a strict linear model with all covariates and the response being normally distributed. \cite{FanSamworthWu:2009} assume a generalized linear model, as does the maximum marginal likelihood estimator (MMLE) method of \cite{MMLE2010}. \cite{XuChen:2014} further explored feature selection in the context of the generalized linear model. Recent publications have proposed feature screening methods that are \textit{non-parametric} or \textit{model-free}, where the assumptions on the underlying model between predictor and response are relaxed or even removed. [See e.g. \cite{FanFengSong:2011}; \cite{zhuLiLiZhu2011}; \cite{DCSIScitation} ; \cite{SuperLongNames}; \cite{CuiLiZhong2015}]. We will further address the distance correlation based method of \cite{DCSIScitation} later in this paper. 
 
Even though these aforementioned methods relax or remove assumptions on the relationship between covariate and response, most SIS-based procedures still tacitly assume that the predictor variables are continuous. 
Notably, this implicit assumption of continuity of the predictors can be limiting, since ultrahigh dimensional data with discrete predictors and discrete responses are rather ubiquitous in practice. 
(For example, the fields of bioinformatics and text mining commonly have need to analyze such data. Gene expression counts in GWAS data is a common example of the first; classifying Chinese text documents by keyword as in \cite{NaiveBayes2014} and \cite{Huang2014} are examples of the latter). This work will specifically focus on the screening of ultrahigh dimensional \textit{categorical} data.

Although there are a number of extant methods for binary (and in some cases multi-class) classification of high dimensional data, including random forests [\cite{breiman2001random}; \cite{randForest2002}], $k$-nearest neighbors [\cite{knn2009}], and support vector machines [\cite{tong2001support}; \cite{kim2005dimension}], these methods become increasingly unstable as the feature space becomes ultrahigh dimensional.

Recognizing the relative dearth of methods for analyzing ultrahigh dimensional categorical data, \cite{Huang2014} presented a method, based on Pearson's Chi-squared Test, for screening categorical data. Hereafter their method will be referred to as HLW-SIS (Huang-Li-Wang-SIS). This deviates from the original name of PC-SIS proposed by Huang et al., however our newly proposed name avoids the similarity with the distance-correlation (DC-SIS) method of \cite{DCSIScitation}.

We propose a new method of screening for data which has both categorical predictor and categorical response values.  Our method has the sure screening property of \cite{FanLv:2008}. Furthermore, under a set of reasonable conditions, we prove that our method correctly identifies the true model consistently, like unto the strong screening property seen in \cite{Huang2014}.
Via simulation, we compare our method to three other methods which admit both categorical predictors and categorical response: MMLE [\cite{MMLE2010}]; DC-SIS [\cite{DCSIScitation}]; and HLW-SIS [\cite{Huang2014}]. We demonstrate that our proposed method has comparable or superior (in some cases, vastly so) screening accuracy for a robust variety of data sets, and moreover requires significantly shorter computation time.    

The rest of this article is organized as follows. In Section \ref{prelims} we describe the premise of the pursuit in question and propose a new screening procedure. In Section 3 we discuss the theoretical properties of our screening method. Section \ref{simulations} contains the details of four simulations using artificially simulated data, as well as the particulars of our method on a real data set from bioinformatics. The results for these simulations and the real data analysis are found in Section \ref{simTables}. 
The final section (Section \ref{proofSection}) is devoted to the proofs of the theoretical results of Section \ref{theory}.




\section{Preliminaries}\label{prelims}
In \cite{Huang2014}, they considered the question of classifying Internet advertisements based on the presence or absence of given keywords. 
They treated each covariate, $X_j$, as binary (although their method allowed for more levels) and the response $Y$ as having $K$-many levels, labeled as $k = 1,2,3, \ldots , K$. Here we treat each \textit{covariate} $X_j$ as having $K_j$-many levels, and assume the response is binary. (So opposite of Huang et al. in a sense). The methods we outline below can be easily extended to a categorical response with greater than two levels, however we will herein only consider binary $Y$. This will allow for some simplification of our notation and proofs. Furthermore, the levels of each covariate can (where appropriate) be taken as being ordinal, so that there is an assumed ordering of the levels:
\[\text{Level } 1 \prec \text{Level } 2 \prec \text{Level } 3 \prec \cdots \prec\text{Level } K_j.\] 
When desired and meaningful, this available premise of level ordinality permits for conclusions pertaining to an exhibited linear trend between the covariates and the response, much like unto the trend test of Cochran [\cite{CochranOrig}] and Armitage [\cite{ArmitageOrig}]. Notably, this enables researchers to form a stronger substantive conclusion about the relationship between the features selected by our  proposed method (see Section \ref{theory}) and the response than was previously available via use of HLW-SIS. In such a case, instead of looking for a general association between the covariates and the response, we can examine and order covariates based on the evidence of a \textit{linear trend} between said covariate and the response. This possibility to examine trend between the response and covariates is, however, only one example of a robust number of settings that our below proposed method is capable of handling. 

Note that we allow for the levels for some or all of the $X_j$'s to be different from the levels of other covariates. Furthermore, we assign a numeric score $v_{k}^{(j)}$ to each level $k$ of $X_j$. Again, when desired and appropriate, the ordering of the $v_{k}^{(j)}$ scores should conform to the ordering of the levels as shown above. For a sequence of $n$ samples of $X_j$, we will denote the (estimated) average level score by $\bar{X}_j$. Since the response $Y$ is considered binary, we will encode its levels using 0 and 1. Then, again for a series of $n$ samples, we will let $\bar{Y} = \frac{1}{n}\sum Y_i$ denote the average response value.

When we need to refer to a general subset of the covariates $X_j$, we will use $X_{i(\mathcal{S})}$, where \[\mathcal{S} \subseteq \{1,2,3,\ldots,p\}\]
is the set of indices for the covariates we wish to discuss. As a matter of simplicity, we will let $\mathcal{S}$ refer to the model consisting of the covariates whose indicies are in $\mathcal{S}$. Define $\mathcal{S}_F = \{1,2,3,\ldots, p\}$ as the full model, which contains all covariates.   
Let $\mathcal{D}\left(Y_i \mid X_{i(\mathcal{S})}\right)$ indicate the conditional distribution of $Y_i$ given $X_{i(\mathcal{S})}$. We will consider a model $\mathcal{S}$ to be sufficient if 
\[\mathcal{D}\left(Y_i \mid X_{i(\mathcal{S}_F)}\right) = \mathcal{D}\left(Y_i \mid X_{i(\mathcal{S})}\right)\]
The full model $\mathcal{S}_F$ is trivially sufficient. We are ultimately only interested in finding the smallest (cardinality-wise) sufficient model. We will call the smallest sufficient model the true model. Our aim in feature screening is to determine an estimated model which contains the true model and is moreover the \textit{smallest} such model to contain the true features.  
The next section will outline the specifics of our proposed screening approach for estimating the true model. As a matter of further notation, we will denote the true model by $\mathcal{S}_T$ and the estimated model by $\widehat{\mathcal{S}}$. 

\section{Using a Cochran-Armitage-like Test Statistic}\label{theory}


The general form for the linear correlation between $X_j$ and $Y$ is given by \[\varrho_j = \frac{\text{cov}(X_j, Y)}{\sigma_j\sigma_Y},\]
where $\text{cov}(X_j, Y)$ is the covariance of $X_j$ versus $Y$, $\sigma_j$ is the standard deviation of $X_j$, and $\sigma_Y$ is the standard deviation of $Y$. 

This brings us to the use of a screening statistic for the purpose of ordering our covariates relative to their estimated correlation with the response.

We will be extending a test statistic outlined by Alan \cite{agrestiBook}, which is directly based on approximating the correlation between $X_j$ and $Y$ when both are categorical. For each $j$ from $1$ to $p$, define the following:
\[\hat{\varrho}_j=\frac{\left|\sum\limits_{\substack{1\leq k \leq K_j \\0 \leq m \leq 1}}(v_k^{(j)}-\bar{v}^{(j)})(m-\bar{Y})\hat{p}_{km}^{(j)}\right|}{\sqrt{\left(\sum\limits_{k = 1}^{K_j}(v_k^{(j)}-\bar{v}^{(j)})^2\hat{p}_k^{(j)}\right)\left(\sum\limits_{m = 0}^1(m-\bar{Y})^2\hat{p}_m\right) }},\]
where $\hat{p}_{km}^{(j)}$, $\hat{p}_{k}^{(j)}$, and $\hat{p}_{m}$ represent the sample estimates (by the relevant sample proportion) of the following probabilities:
\[{p}_{km}^{(j)} = \mathbb{P}(X_j = k, Y = m), \quad{p}_{k}^{(j)} = \mathbb{P}(X_j = k),\quad {p}_m = \mathbb{P}(Y = m).\]

(As can already be seen, the notation for this can become \textit{exceedingly} messy). Note that $\rhoHat$ has been constructed to be non-negative.
A simpler version of $\rhoHat$ (given without the indexing by $j$) is presented in \cite{agrestiBook} as a generalization of the Cochran-Armitage test for trend. As discussed previously, in the proper setting, our method can be specifically interpreted as screening for the covariates which exhibit the strongest linear trend in relation to the response.  
It should be noted here that this newly proposed method establishes a generalization of the Pearson correlation based method of \cite{FanLv:2008}. While they assume that all predictors and the response are spherically distributed random variables, we assume no specific distribution for the covariates or the response. While our main focus herein is on categorical data, Simulation 4 in Section \ref{simulations} suggests that the Pearson correlation can be effectively used on continuous data in broader settings than originally allowed by \cite{FanLv:2008}.  

Using the $\rhoHat$, we form the estimated model $\widehat{\mathcal{S}}$ by selecting a cutoff $c >0$. Define $\widehat{\mathcal{S}}$ as follows:
\[\widehat{\mathcal{S}} = \{j: 1\leq j \leq p, ~ \rhoHat > c\}.\]
Let the numerator of $\rhoHat$ be designated by $\hat{\tau}_j$. Note that the denominator of $\rhoHat$ consists of (biased) sample estimators for the standard deviations of $X_j$ and $Y$. (However, the bias of these estimators disappears asymptotically). Both of these estimators are consistent estimators of their respective standard deviations. Consistency is easy to prove using Chebychev's inequality and routine algebra. For completeness, this will be shown shortly herein.

\subsection{Theoretical properties}
We now define two conditions:
\begin{enumerate}
    \item[(C1)] \emph{Bounds on the standard deviations}. Assume that there exists a positive constant $\sigma_{\text{min}}$ such that for all $j$,\[\sigma_j > \sigma_{\text{min}}\quad\quad\text{and}\quad\quad\sigma_Y > \sigma_{\text{min}}\]
    This excludes features that are constant and hence have a standard deviation of 0. 
    It should further be noted that a sufficient upper bound on $\sigma_j$ and $\sigma_Y$ can also be obtained, by use of Popoviciu's inequality on variances [see \cite{Popoviciu}]:
\[\text{Let } \quad \sigma_{\text{max}} = \text{max}\left\{\frac{1}{2},~ \sqrt{\frac{1}{4}\left(v_{K_j}^{(j)}-{v}_{1}^{(j)}\right)}\right\},\]
where the first term in the maximum selection is a bound on the standard deviation of $Y$ and the second term is given by Popoviciu's inequality on variances. This $\sigma_{\text{max}}$ acts as an upper bound for both $\sigma_j$ and $\sigma_Y$ simultaneously. 
    

    
    \item[(C2)] \textit{Marginal Covariances}. Assume that $\varrho_j = 0$ for any $j \not\in \mathcal{S}_T$. Define
    \[\omega_{km}^{(j)} = \left| (v_k^{(j)}-\mathbb{E}(X_j))(m-\mathbb{E}(Y)){p}_{km}^{(j)} \right| .\]
    Assume there exists a positive constant $\omega_{\text{min}}$ such that \[\min_{j \in \mathcal{S}_T}\left(\max_{\substack{1\leq k \leq K_j \\0 \leq m \leq 1}}\left\{\omega_{km}^{(j)}\right\}\right) > \omega_{\min}>0\]
    
    This places a lower bound on the smallest (indexing by $j$) of the maximum values of the $\omega_{km}^{(j)}$.
    Note that (C2) requires that for every true feature (i.e. $j \in \mathcal{S}_T$), there exists at least one level of the response $Y$ and one level of the feature $X_j$ that are marginally correlated (i.e. $\omega_{km}^{(j)} > \omega_{\min}$). 
    This is of course a natural assumption to make for the true features and should be quite easy to satisfy in a wide variety of reasonable situations.
    
    
\end{enumerate}

This brings us to the following theorems:

\subsubsection{Theorem 1}\label{Thm1} (\textit{Strong Screening Consistency}). Given conditions (C1) and (C2), there exists a positive constant $c>0$ such that \[\mathbb{P}(\widehat{\mathcal{S}} = \mathcal{S}_T) \longrightarrow 1 \text{ as } n \longrightarrow \infty.\]

\subsubsection{Theorem 2}\label{Thm2} (\textit{Weak Screening Consistency}). Given  that conditions (C1) still holds, while removing from (C2) only the assumption of $\varrho_j = 0$ for all $j \notin \mathcal{S}_T$, there exists a positive constant $c>0$ such that \[\mathbb{P}(\widehat{\mathcal{S}} \supseteq \mathcal{S}_T) \longrightarrow 1 \text{ as } n \longrightarrow \infty.\]
(But $\mathbb{P}(\widehat{\mathcal{S}} \subseteq \mathcal{S}_T)$ may not converge to 1 as $n$ approaches infinity). 

The proofs of these two theorems are presented in Section \ref{proofSection}.

\subsection{Corollaries}
We can draw several corollaries from the proofs of Theorems 1 and 2 (see Section \ref{proofSection}). These results are not themselves about sure screening, but they are nevertheless important observations on the underlying mechanics of our method. 

\subsubsection{Corollary 1}\label{Cor1}
In Step 1 of the proofs of Theorems 1 and 2, it will be shown that there exists a value $\varrho_{\min}$ such that for any $j \in \mathcal{S}_T$, we have $\varrho_j > \varrho_{\min}$. 

\subsubsection{Corollary 2}\label{Cor2}
From the end of Step 2 in the proofs of Theorems 1 and 2, we will conclude that $\rhoHat$ converges \textit{uniformly} in probability  to $\varrho_j$. In other words, 
\[\mathbb{P}\left(\max_{1 \leq j \leq p}|\rhoHat - \varrho_j| > \varepsilon\right) \rightarrow 0 \quad \text{ as } n\rightarrow\infty\] for any $\varepsilon> 0$.

\subsection{Comments on Choosing a Sufficient Cutoff}
Although we will show (in the proof of Theorem 1) that a constant $c$ exists such that
\[\widehat{\mathcal{S}} = \{j: 1\leq j \leq p, ~ \rhoHat > c\}\] converges with probability 1 to $\mathcal{S}_T$, we have yet to discuss a method for actually determining such cutoff.
An equivalent problem is that of determining a positive integer $d_0$ such that if we let 
\[\widehat{\mathcal{S}} = \{j : 1\leq j \leq p \text{ and } \rhoHat \text{ is one of the $d_0$ largest $\rhoHatSub{}$} \}\]
\cite{Huang2014} present a possible approach for determining an estimate for such a $d_0$ using the ratio of adjacent (when ordered from greatest to least) screening statistics. They argue that if we order the screening statistics from largest to smallest
\[\rhoHatSub{(1)} \geq \rhoHatSub{(2)} \geq \cdots \geq \rhoHatSub{(p)}\]
(where $\rhoHatSub{(k)}$ is the $k$th largest screening statistic), then we can estimate $d_0$ by
\[\hat{d} = \text{argmax}_{1 \leq j \leq p} ~\left\{ \frac{\rhoHatSub{(j)}}{\rhoHatSub{(j + 1)}}\right\}.\]

This estimation comes from the fact that if $ j = d_0$, then $\rhoHatSub{(j)} > \varrho_{\min} >0$ (see Corollary 1 at \ref{Cor1}), but $\rhoHatSub{(j+1)}\xrightarrow{p} 0$. From a theoretical perspective, this will in turn imply that we have $\rhoHatSub{(j)}/\rhoHatSub{(j + 1)}\xrightarrow{p} \infty$. 
However, implementing this method in practice can be challenging, 
since care must be taken to not select covariates associated with minuscule $\rhoHatSub{}$, yet which at the same time have a relatively large ratio between it and the next smallest $\rhoHatSub{}$. For example, if we have three covariates $X_1$, $X_2$, and $X_3$ to select from and their respective screening statistics are \[\rhoHatSub{1} = 0.8, \quad \rhoHatSub{2} = 0.00008, \quad \rhoHatSub{3} = 8\times 10^{-10},\]
we can see explicitly that $X_2$ and $X_3$ likely have almost no causative effect on the response. Yet, if we apply the above suggested method for estimating $d_0$, both features $X_1$ and $X_2$ will be selected as relevant. While here only one covariate beyond what we would intuitively expect to be the true model was selected, in the presence of thousands (or even millions) of possible predictors, such overestimation of $d_0$ can prove non-trivial. (Picture for example 25 covariates with $\rhoHat = 0.8$, 2500 covariates with $\rhoHat = 0.00008$ and one covariate with $\rhoHat = 8 \times 10^{-10}$).  

One heuristic fix that we attempted was dropping all screening statistics below various cutoff levels (e.g. $10^{-5}, 10^{-6},$ etc.). Importantly, however, note that this brings us  back philosophically to the same question of selecting a sufficient cutoff for which features to retain. Few papers currently exist on the topic of deterministically approximating the true model size (the ideal $d_0$). \cite{KongWangWahba2015} present one possible approach in the setting of DC-SIS. 
Overall, we suggest that readers proceed with caution when trying to adaptively determine a cutoff for our proposed method.     

\section{Simulations and Empirical Data Analysis}\label{simulations}
We performed four simulations on artificially generated data to validate our theoretical results empirically. Each of these simulations, as well as the associated results, are summarized below. (See Section \ref{simTables} for the results). We also performed an analysis on an empirical data set from the NCBI databases examining polycystic ovary syndrome (PCOS).  

\subsection{Simulation 1} 
In this simulation, we will be observing $200$ samples ($n = 200$) of $5000$ covaraiates ($p = 5000$). Of these $p$-many covariates, only 10 of them ($X_1, ~X_2, ~X_3, \ldots, X_{10}$) will be constructed to have meaningful contribution to the outcome $Y$. These covariates will be referred to as the causative predictors. Our goal is to examine the minimum model size for which all of the causative covariates will be included. We will run 500 replications and record the minimum model size required for each replication. The test data is the same for all four methods examined herein (our method, MMLE, DC-SIS, and HLW-SIS).  

The $Y_i$ are generated by a Bernoulli process with $\PP(Y = 1) = p_y,$ where $p_y \sim \text{unif}(0.05, 0.95)$ is chosen anew for each replicate of the simulation.  

The covariates $X_j$ will take on values of 0, 1, or 2 (representative of three ordinal levels, with $0 \prec 1 \prec 2$). For $1 \leq j \leq 10$, let 
\[\mathbb{P}(X_{ij} = k \mid Y_i = m) = \theta_{mk}\] be determined by the binomial distribution of the number of successes over two independent Bernoulli trials each with probability $\pi_{mj}$ as given below in Table 1.

\begin{table}
\caption{Values of $\pi_{mj}$}
\begin{center}
\begin{tabular}{ |c | c  c c c c c c c c c| }
\hline
& $\pi_{m1}$ &$\pi_{m2}$ &$\pi_{m3}$ &$\pi_{m4}$ &$\pi_{m5}$ &$\pi_{m6}$ &$\pi_{m7}$ &$\pi_{m8}$ &$\pi_{m9}$ &$\pi_{m,10}$ \\
  \hline\hline			
  $Y = 0$ & 0.3 &0.4  &0.6 &0.7 &0.2 &0.4 &0.3 &0.8 &0.4 &0.2\\\hline
  $Y =1$ & 0.6 &0.1  &0.1 &0.4 &0.8 &0.7 &0.9 &0.2 &0.7 &0.6\\
  \hline  
\end{tabular}

\end{center}
\end{table}

To wit, since $\pi_{mj}$ represents the probability of ``success" in the Bernoulli trials used to determine the value of $X_{ij}$ when $Y_i = m$, then 
\[\theta_{mk} = \binom{2}{k}\pi_{mj}^k(1-\pi_{mj})^{2-k}.\]

For $j > 10$, let $X_j \sim \text{binomial}(2, p_j)$, where $p_j \sim \text{Unif}(0.05, 0.95)$ is chosen for each j. Thus the sampling of these covariates is done without respect to the value of $Y_i$.  
As with the generation of the $Y$s, $p_j$ is chosen anew for each replication of the simulation. 

This use of the binomial distribution to determine the value of each $X_{ij}$ is of importance to genetic applications in that it in many ways models the pairing of dominant and recessive alleles, with varying degrees of probability of a dominant allele being present. 

This can be elucidated as follows: Let $D$ be a Bernoulli random variable with probability of ``success" ($D = 1$) being $\pi$. We then can assign dominant or recessive alleles to the support of $D$:
\[0 \longrightarrow a \quad\quad\quad 1 \longrightarrow A \]
In this way, if we examine two identical but independent trials of $D$, we can form genotypes $aa$, $aA = Aa$, and $AA$. Based on the probability $\pi$, we can determine the probability of each genotype occurring. The former probability ($\pi$) is equivalent to $\pi_{mj}$ above. The latter probability is equivalent to $\theta_{mk}$. Hence the levels of $X_j$ can be taken as representing possible genotypes.

The recorded outcomes of our simulations are two-fold: We first report the mean minimum model size over the 500 replications of the simulation. This refers to the average number of covariates that needed to be selected to contain the true causative predictors. We also record the proportion (out of 500 replications) of screening acquisition of each causative covariate individually for model sizes 10, 15, and 20. This can be taken as the power with which we correctly select each covariate in $\mathcal{S}_T$ when $\widehat{S}$ consists of the covariates associated with the 10 highest, the 15 highest, and the 20 highest screening scores. These aforementioned results for Simulation 1 are summarized in Section \ref{simTables}.

\subsection{Simulation 2}
Simulation 2 is formulated to establish the superior ability of the trend test method to screen and select covariates which are linearly correlated with the response. This simulation bears a resemblance to Example 3 of \cite{Huang2014} in that it involves discretizing a normally distributed continuous variable in order to view it in a categorical setting. As with Simulation 1, we examine 200 samples of 5000 total covariates; moreover we again take the first 10 covariates as the causative features that we wish to select. This simulation is replicated 500 times. The test data is the same for all four methods examined herein.

The 200 samples of the response, $Y$, are created first. This is accomplished by the same methods of Simulation 1: the $Y_i$ are generated by a Bernoulli process with $\PP(Y = 1) = p_y,$ where $p_y \sim \text{unif}(0.05, 0.95)$ is chosen anew for each replicate of the simulation. After generating the 200 samplings of $Y$, we generate the corresponding 200 samplings of each of the 5000 covariates. The non-causative covariates are created using an approach identical to that of Simulation 1. Again, this is done with no regard to the value of the associated $Y_i.$

The causative covariates (viz. $X_1$ through $X_{10}$) are generated as follows: Given $Y_i$, we take a random sample from the normal distribution with mean equal to $Y_i$ (either 0 or 1) and standard deviation equal to 1. Call the value obtained from this sampling $Z_{ij}$. We then create $X_{ij}$ based on the cutoffs $(\kappa_{Lj}, \kappa_{Uj})$ listed in Table \ref{CutoffTable} and the following criterion:
\[X_{ij} = \begin{cases}
0 & \mbox{if } Z_{ij} < \kappa_{Lj}\\
1 & \mbox{if } \kappa_{Lj} \leq  Z_{ij} \leq \kappa_{Uj}\\
2 & \mbox{if } \kappa_{Uj} < Z_{ij}
\end{cases}\]

\begin{table}
\caption{Cutoff Values}
\begin{center}
\label{CutoffTable}
\begin{tabular}{ |c | c  c c c c c c c c c| }
\hline
& $X_1$ &$X_2$ &$X_3$ &$X_4$ &$X_5$ &$X_6$ &$X_7$ &$X_8$ &$X_9$ &$X_{10}$ \\
  \hline\hline			
  $\kappa_{Lj}$ &0 &0  &0.2 &0 &-0.213 &0.25 &0 &0.1 &-0.2 &0.213\\\hline
  $\kappa_{Uj}$& 0.75 &1 &0.8 &0.9 &1.213&1&1  &1 &1.2 &0.787 \\
  \hline  
\end{tabular}

\end{center}
\end{table}

This process creates causative covariates which are pairwise linearly correlated with the response. Since each of the four methods are highly accurate ($\geq$ 99\%) in correctly identify the causative feature when their correlation with $Y$ is moderately high (e.g. Pearson correlation coefficient $r \geq$ 0.7), we have selected cutoff pairs that lead to a Pearson correlation coefficient between 0.25 and 0.65 for $Y$ pairwise with each of the first ten covariates. (This is a heuristic, not absolute, range). It should be noted that while all causative predictors are constructed to have positive linear correlation with $Y$, covariates with negative correlation yield identical results. This is due to the fact that we only care about the magnitude of $\hat{\varrho_j}$.  


\subsection{Simulation 3} This simulation is meant to resemble data based on a form of logistic regression, which is a strength of MMLE. Nevertheless, we will see that our method performs admirably in this setting and produces results abreast with that of MMLE. As an aside, it is necessary to note that, since MMLE requires solving an optimization problem to produce its results, our method will be significantly faster to run.

The simulation data for Simulation 3 is created as follows. First generate $n = 200$ samplings of each $X_j$ ($1 \leq j \leq p$, with $p = 5000$) by uniformly sampling the set $\{0,1,2\}$ with equal probability. Then calculate

\[L = \sum_{j = 1}^5 [I(X_j = 0) \times \theta_{X_{j}=0} + I(X_j = 1) \times \theta_{X_{j}=1} + I(X_j = 2) \times \theta_{X_{j}=2}],\]

where each $\theta_{X_j = k}$ is given in Table \ref{thetaTable}. Note that here we are only taking the first five covariates ($X_1$ through $X_5$) as causative. 

\begin{table}
\begin{center}
\caption{Coefficients for $L$}
\begin{tabular}{|l|ccccc|}
\hline
 & $\theta_{X_{1}}$ & $\theta_{X_{2}}$ & $\theta_{X_{3}}$ & $\theta_{X_{4}}$ & $\theta_{X_{5}}$ \\ \hline\hline
$X_k = 0$ & 0 & -5 & 2 & -6 & 1 \\
$X_k = 1$ & 3 & -3 & 4 & -4 & 3 \\ 
$X_k = 2$ & 5 & -1 & 6 & -2 & 5 \\ \hline
\end{tabular}
\label{thetaTable}
\end{center}
\end{table}

Now generate each $Y_i$ as a Bernoulli process with
\begin{center}
$P(Y = 1) = \frac{1}{1 + \text{exp}(-L)}.$
\end{center}

We perform 500 replication of this simulation, with each of the four methods being examined under the same data sets. The results of Simulation 3 are given in Section \ref{simTables}.

\subsection{Simulation 4}
Although our method is not originally designed or emphasized for use on continuous data, this simulation presents a comparison or our method versus DC-SIS when the covariates are normally distributed. 
The motivation for this simulation is the statement by \cite{DCSIScitation} that when the covariates are normally distributed, DC-SIS is ``equivalent (although not equal, see Theorem 7 of \cite{szekely2007}) to the method of \cite{FanLv:2008}. However, \cite{szekely2007} and \cite{szekely2009} further elucidate the fact that the response must also be normally distributed for DC-SIS to be equivalent to Pearson correlation. Our aim here is to see how DC-SIS performs when the covaraites are normally distributed, yet the response is not necessarily normally distributed. 



The data for this simulation is generated as follows. We will observe 200 samplings ($n = 200$) of 1000 covariates ($p = 1000$). 
Let $\mathbf{X_i}$ be a vector of length 1000, where $\mathbf{X_i} \sim \text{MVN}(\textbf{0}, \Sigma)$ is sampled for $i$ from 1 to 200. Here the covariance matrix $\Sigma = [\sigma_{j_1j_2}]$ is given by $\sigma_{j_1j_2} = 0.2^{|j_1-j_2|}$.  Now we generate the response $Y$ using only the first 10 predictors. 
Specifically, we let $Y_i = \mathbf{X_i}\bm{\beta}$, where $\bm{\beta}$ is a vector of length 10 defined in Table \ref{betas}. Note that, because the first ten $X_j$s are not independent, this construction of $Y$ does not guarantee that $Y$ itself is normally distributed.  

\begin{table}
\begin{center}
\caption{Defining $\bm{\beta}$}
\begin{tabular}{|cccccccccc|}
\hline
$\beta_1$ & $\beta_2$ & $\beta_3$ & $\beta_4$ & $\beta_5$ & $\beta_6$ & $\beta_7$ & $\beta_8$ & $\beta_9$ & $\beta_{10}$ \\ \hline\hline
5 & -5 & 5.5 & -6 & 6 & 4 & 4.5 & -5.5 & 5 & -4 \\ \hline
\end{tabular}
\label{betas}
\end{center}
\end{table}

We perform 500 replication of this simulation, with both methods (ours and DC-SIS) being examined under the same data sets. The results of Simulation 4 are given in Section \ref{simTables}.

\subsection{Comments on Simulation Results}
In Simulations 1, and 2, our method results in the smallest average model size required to contain the true model. Since these simulations were designed to specifically take advantage of the relation of our method to a test for linear trend, these results should not be surprising. The results for MMLE in these first two simulations are less than inspiring.  The overall results of these first two simulations suggests that our method is more robust in the presence of data with an unbalanced amount of positive ($Y = 1$) responses. 

In Simulation 3, we once again obtain a smaller required mean model size than DC-SIS and HLW-SIS. 
In the case of HLW-SIS, our method obtains noticeably smaller required minimum model sizes. These results in comparison to the mean model sizes for HLW-SIS are appealing since HLW-SIS was  presented as a worthy method for screening ultra-high dimensional feature spaces. A specific comment on the results of Simulation 3 for MMLE-SIS is in order. As was previously discussed, a strength of MMLE-SIS is screening data in a logistic regression setting. Indeed, MMLE recoups its earlier collapses and matches (by about four hundredths of an average minimum model size) our method nearly perfectly. However, as has been previously noted, since MMLE requires solving an optimization problem to produce its screening statistics, our newly proposed method is significantly faster in computational run time. Thus, when run time is an issue, we suggest the use of our method over MMLE even in a logistic regression setting.

For Simulation 4, we obtain the largest gap (of the four simulations considered) in mean minimum model size between our method and DC-SIS. This suggests that, under the conditions prescribed by Simulation 4, the generalization of Pearson correlation to continuous, but not necessarily normally distributed, data may prove superior to extant methods such as DC-SIS.    

\subsection{Real Data Analysis}
We apply a two stage iterative process to a clinical data set examining polycystic ovary syndrome (PCOS). Following a strict approval process, this PCOS data was downloaded from the database of genotypes and phenotypes (dbGaP) of the National Center for Biotechnology Information (NCBI) at the NIH (dbGaP Study Accession: \href{https://www.ncbi.nlm.nih.gov/projects/gap/cgi-bin/study.cgi?study_id=phs000368.v1.p1}{phs000368.v1.p1}). This data consists of 4099 observation (1043 cases, 3056 controls) of each of 731,442 SNPs. The response is PCOS affection status and the predictors are the encoded SNP geneotype values. Our real data analysis is modeled after that of \cite{ZhongZhu2014} and \cite{LiuLiWu2014}. Specifically, using the iterative screening approach outlined in \cite{ZhongZhu2014} for their DC-ISIS procedure, we first iterate over the values $p_1 = 5,6,\ldots, [n/\log(n)] = 493$ to determine a value for $p_1$. The optimal value for $p_1$ is that which minimizes the MSPE for logistic regression over the remaining $p_2 = [n/\log(n)]-p_1$ predictors in question over 200 random replications each time using 75\% of the data for training and 25\% for testing. We found that $p_1 = 191$ and $p_2 = [n/\log(n)]-p_1 = 302$ as initial values minimized the MSPE in our case.

After screening the real data set using the iterative application of our proposed method, we obtain a relatively small set of SNPs with positive screening scores scores (450 such SNPs). Following the process of \cite{LiuLiWu2014}, we select a submodel with size $d = [n^{4/5}/\log\left(n^{4/5}\right)] = 117$, where the SNPs corresponding to the $d$ largest iterative screening scores are chosen.  

Using 10-fold cross validation in the R package \texttt{glmnet}, we then post screen our selected $d$ many SNPs via a variety of penalized regression methods to further reduce the final model size. We use three such techniques: lasso [\cite{Tibshirani}], adaptive-lasso [\cite{ZouAdaptLasso2006}], and elastic net (with $\alpha = 0.09$; see below for the use of $\alpha$) [\cite{elasticNet}]. 
Each of these three methods employs penalized logistic regression of the negative binomial log-likelihood, which is as follows:
\[\min_{\bm{\beta}\in \mathbb{R}^{p}}\left\{ -\left[\frac{1}{N}\sum_{i=1}^N y_i(x^T_i\bm{\beta})-\log(1+e^{x^T_i\bm{\beta}})\right]+\lambda\left[\frac{(1-\alpha)}{2}\lVert \bm{\beta}\rVert_2^2+\alpha\lVert\bm{\beta}\rVert_1\right]\right\}.\]

The aggressiveness of the penalty is controlled by a parameter $\lambda$. The parameter $\lambda$ is chosen using a cross-validated coordinate descent approach, where the objective is minimizing the predicted misclassification rate. This process is handled internally in the \texttt{glmnet} package in R [\cite{glmnetCite}]. 
When $\alpha = 1$ above, we have the lasso penalty function. To perform adaptive lasso, we first fit weights for each component of $\bm{\beta}$ using ridge regression ($\alpha = 0$). These weights are then enforced in \texttt{glmnet} by use of the \texttt{penalty.factor} option while preforming lasso. 
Our elastic net model is tuned in a manner similar to the original \cite{elasticNet} paper. 
We first pick a grid of values for $\alpha$. For simplicity we used $\alpha_k = \left\{\frac{k}{100}\right\}$ for $k = 1,2,3,\ldots 99$. (When $\alpha = 1$, this is lasso, which is examined separately above). Then, for each $\alpha_k$, we fit a model for our $d$ many parameters using elastic net. As with lasso and adaptive lasso, the other tuning parameter, $\lambda$, is selected by tenfold CV. The
chosen $\lambda$ is the one giving the smallest 10-fold cross validated misclassification error. Here, our tuning procedures found $\alpha = 0.09$ to be the $\alpha$ for which misclassification error was minimized. 

The empirical results of our final model selection process are summarized in Table \ref{realDataSummary}:

\begin{table}
\centering
\caption{Empirical results of real data analysis.}
\begin{tabular}{|c| c c c c|} 
 \hline
  Post Screening Method& Model size & McFadden's pseudo-$R^2$ & AIC& Misclass. rate \\ [0.5ex] 
 \hline\hline 
 Lasso & 71 & 0.1784 &-1735.27& 21.59\% \\ 
 Adaptive Lasso &56  & 0.1761&-1755.10& 21.08\% \\
 Elastic Net ($\alpha = 0.09$)& 91 & 0.1799 &-1691.66& 21.15\% \\ [1ex] 
 \hline
\end{tabular}
\label{realDataSummary}
\end{table}

As a measure for goodness-of-fit, we include the McFadden's pseudo-$R^2$ value in the table [see \cite{McFadden1974}]. For further justification for the use of McFadden's pseudo-$R^2$ see \cite{menard2000}.
It should be noted that McFadden's pseudo-$R^2$ does not have an intuitive interpretation like unto Pearson's traditional $R^2$. In \cite{louviere2000stated}, McFadden suggests that a model having a pseudo-$R^2$ even in the range of $0.2$ to $0.4$ can be taken as having excellent fit [see also \cite{DomencichMcFadden1975}]. From this, we conclude that our four fitted models above all have sufficient fit. Based on the relative parsimony of the adaptive lasso model, as well as its comparatively similar pseudo-$R^2$ and misclassification rate to the other methods, we suggest the use of the model found by adpative lasso as the final model. This suggestion is supported by comparing the Akaike's Information Criterion (AIC) of each model, the minimal AIC being that associated with the adaptive lasso model.

\section{Results of Simulations}\label{simTables}
Here we present the results of our four simulations. In each table of results, our newly proposed method is referred to by the working title of CAT-SIS (Categorical-SIS).  
\subsection{Simulation 1 Results}

The results of Simulation 1 are summarized in Tables 6 through 10:

\begin{table}
\caption{Mean Minimum Model Sizes $(n = 200,~ p = 5000)$}
\begin{center}
\begin{tabular}{ |c | c c c c | }
\hline
 & CAT-SIS & MMLE & DC-SIS & HLW-SIS \\ \hline\hline
Mean Minimum Model Size & 54.674 & 150.340 & 64.990 & 93.018 \\ 
  \hline  
\end{tabular}
\end{center}
\end{table}

\begin{table}
\begin{center}
\caption{Proportion of Replications Where $X_j$ is in the Top $d$ Causative Covariates}
\begin{tabular}{|l|cccccccccc|}
\hline
\multicolumn{11}{|c|}{CAT-SIS} \\ \hline
 & $X_{1}$ & $X_{2}$ & $X_{3}$ & $X_{4}$ & $X_{5}$ & $X_{6}$ & $X_{7}$ & $X_{8}$ & $X_{9}$ & $X_{10}$ \\ \hline\hline
$d = 10$ & 0.864 & 0.980 & 0.988 & 0.832 & 1.000 & 0.998 & 0.844 & 0.974 & 0.836 & 0.858 \\
$d = 15$ & 0.916 & 0.988 & 0.994 & 0.922 & 1.000 & 0.998 & 0.912 & 0.982 & 0.904 & 0.908 \\
$d = 20$ & 0.920 & 0.988 & 0.994 & 0.930 & 1.000 & 0.998 & 0.926 & 0.990 & 0.930 & 0.930 \\ \hline
\end{tabular}
\end{center}
\end{table}

\begin{table}
\begin{center}
\caption{Proportion of Replications Where $X_j$ is in the Top $d$ Causative Covariates}
\begin{tabular}{|l|cccccccccc|}
\hline
\multicolumn{11}{|c|}{MMLE} \\ \hline
 & $X_{1}$ & $X_{2}$ & $X_{3}$ & $X_{4}$ & $X_{5}$ & $X_{6}$ & $X_{7}$ & $X_{8}$ & $X_{9}$ & $X_{10}$ \\ \hline\hline
$d = 10$ & 0.210 & 0.680 & 0.690 & 0.250 & 0.786 & 0.816 & 0.194 & 0.646 & 0.182 & 0.218 \\
$d = 15$ & 0.384 & 0.746 & 0.756 & 0.404 & 0.822 & 0.844 & 0.354 & 0.742 & 0.320 & 0.400 \\
$d = 20$ & 0.508 & 0.796 & 0.800 & 0.490 & 0.854 & 0.864 & 0.452 & 0.796 & 0.440 & 0.490 \\ \hline
\end{tabular}
\end{center}
\end{table}

\begin{table}
\begin{center}
\caption{Proportion of Replications Where $X_j$ is in the Top $d$ Causative Covariates}
\begin{tabular}{|l|cccccccccc|}
\hline
\multicolumn{11}{|c|}{DC-SIS} \\ \hline
 & $X_{1}$ & $X_{2}$ & $X_{3}$ & $X_{4}$ & $X_{5}$ & $X_{6}$ & $X_{7}$ & $X_{8}$ & $X_{9}$ & $X_{10}$ \\ \hline\hline
$d = 10$ & 0.850 & 0.978 & 0.982 & 0.818 & 0.998 & 0.998 & 0.804 & 0.968 & 0.824 & 0.834 \\
$d = 15$ & 0.900 & 0.984 & 0.990 & 0.898 & 0.998 & 0.998 & 0.894 & 0.984 & 0.888 & 0.894 \\
$d = 20$ & 0.916 & 0.988 & 0.992 & 0.920 & 0.998 & 0.998 & 0.922 & 0.986 & 0.908 & 0.914 \\ \hline
\end{tabular}
\end{center}
\end{table}

\begin{table}
\begin{center}
\caption{Proportion of Replications Where $X_j$ is in the Top $d$ Causative Covariates}
\begin{tabular}{|l|cccccccccc|}
\hline
\multicolumn{11}{|c|}{HLW-SIS} \\ \hline
 & $X_{1}$ & $X_{2}$ & $X_{3}$ & $X_{4}$ & $X_{5}$ & $X_{6}$ & $X_{7}$ & $X_{8}$ & $X_{9}$ & $X_{10}$ \\ \hline\hline
$d = 10$ & 0.808 & 0.956 & 0.962 & 0.800 & 0.992 & 0.996 & 0.778 & 0.954 & 0.760 & 0.802 \\
$d = 15$ & 0.862 & 0.974 & 0.980 & 0.864 & 0.994 & 0.998 & 0.860 & 0.966 & 0.854 & 0.862 \\
$d = 20$ & 0.890 & 0.980 & 0.986 & 0.886 & 0.994 & 0.998 & 0.872 & 0.972 & 0.886 & 0.882 \\ \hline
\end{tabular}
\end{center}
\end{table}

\subsection{Simulation 2 Results}
The results of Simulation 2 are summarized in Tables 11 through 15:
\begin{table}
\caption{Mean Minimum Model Sizes $(n = 200,~ p = 5000)$}
\begin{center}
\begin{tabular}{ |c | c c c c | }
\hline
 & CAT-SIS & MMLE & DC-SIS & HLW-SIS \\ \hline\hline
Mean Minimum Model Size & 112.627 & 508.672 & 125.258 & 171.829 \\ 
  \hline  
\end{tabular}
\end{center}
\end{table}

\begin{table}
\begin{center}
\caption{Proportion of Replications Where $X_j$ is in the Top $d$ Causative Covariates}
\begin{tabular}{|l|cccccccccc|}
\hline
\multicolumn{11}{|c|}{CAT-SIS} \\ \hline
 & $X_{1}$ & $X_{2}$ & $X_{3}$ & $X_{4}$ & $X_{5}$ & $X_{6}$ & $X_{7}$ & $X_{8}$ & $X_{9}$ & $X_{10}$ \\ \hline\hline
$d = 10$ & 0.828 & 0.822 & 0.816 & 0.866 & 0.884 & 0.820 & 0.892 & 0.850 & 0.796 & 0.862 \\
$d = 15$ & 0.876 & 0.876 & 0.882 & 0.904 & 0.924 & 0.874 & 0.920 & 0.906 & 0.878 & 0.900 \\
$d = 20$ & 0.888 & 0.892 & 0.898 & 0.918 & 0.940 & 0.902 & 0.936 & 0.920 & 0.898 & 0.920 \\ \hline
\end{tabular}
\end{center}
\end{table}

\begin{table}
\begin{center}
\caption{Proportion of Replications Where $X_j$ is in the Top $d$ Causative Covariates}
\begin{tabular}{|l|cccccccccc|}
\hline
\multicolumn{11}{|c|}{MMLE} \\ \hline
 & $X_{1}$ & $X_{2}$ & $X_{3}$ & $X_{4}$ & $X_{5}$ & $X_{6}$ & $X_{7}$ & $X_{8}$ & $X_{9}$ & $X_{10}$ \\ \hline\hline
$d = 10$ & 0.028 & 0.026 & 0.038 & 0.038 & 0.424 & 0.016 & 0.302 & 0.054 & 0.014 & 0.124 \\
$d = 15$ & 0.072 & 0.060 & 0.066 & 0.078 & 0.554 & 0.054 & 0.388 & 0.098 & 0.032 & 0.204 \\
$d = 20$ & 0.110 & 0.098 & 0.106 & 0.152 & 0.612 & 0.086 & 0.468 & 0.166 & 0.064 & 0.270 \\ \hline
\end{tabular}
\end{center}
\end{table}

\begin{table}
\begin{center}
\caption{Proportion of Replications Where $X_j$ is in the Top $d$ Causative Covariates}
\begin{tabular}{|l|cccccccccc|}
\hline
\multicolumn{11}{|c|}{DC-SIS} \\ \hline
 & $X_{1}$ & $X_{2}$ & $X_{3}$ & $X_{4}$ & $X_{5}$ & $X_{6}$ & $X_{7}$ & $X_{8}$ & $X_{9}$ & $X_{10}$ \\ \hline\hline
$d = 10$ & 0.832 & 0.832 & 0.830 & 0.868 & 0.860 & 0.824 & 0.880 & 0.860 & 0.818 & 0.868 \\
$d = 15$ & 0.876 & 0.884 & 0.886 & 0.904 & 0.886 & 0.880 & 0.880 & 0.910 & 0.878 & 0.906 \\
$d = 20$ & 0.898 & 0.900 & 0.908 & 0.918 & 0.918 & 0.904 & 0.930 & 0.918 & 0.896 & 0.930 \\ \hline
\end{tabular}
\end{center}
\end{table}

\begin{table}
\begin{center}
\caption{Proportion of Replications Where $X_j$ is in the Top $d$ Causative Covariates}
\begin{tabular}{|l|cccccccccc|}
\hline
\multicolumn{11}{|c|}{HLW-SIS} \\ \hline
 & $X_{1}$ & $X_{2}$ & $X_{3}$ & $X_{4}$ & $X_{5}$ & $X_{6}$ & $X_{7}$ & $X_{8}$ & $X_{9}$ & $X_{10}$ \\ \hline\hline
$d = 10$ & 0.754 & 0.766 & 0.744 & 0.774 & 0.846 & 0.752 & 0.844 & 0.776 & 0.738 & 0.808 \\
$d = 15$ & 0.820 & 0.806 & 0.816 & 0.842 & 0.876 & 0.806 & 0.876 & 0.830 & 0.820 & 0.858 \\
$d = 20$ & 0.844 & 0.830 & 0.850 & 0.862 & 0.880 & 0.824 & 0.888 & 0.862 & 0.838 & 0.876 \\ \hline
\end{tabular}
\end{center}
\end{table}

\subsection{Simulation 3 Results}
The results of Simulation 3 are summarized in Tables 16 through 20:
\begin{table}
\caption{Mean Minimum Model Sizes $(n = 200,~ p = 5000)$}
\begin{center}
\begin{tabular}{ |c | c c c c | }
\hline
 & CAT-SIS &  DC-SIS& MMLE & HLW-SIS \\ \hline\hline
Mean Minimum Model Size & 41.976 &46.470 & 41.934  & 93.270 \\ 
  \hline  
\end{tabular}
\end{center}
\end{table}

\begin{table}
\begin{center}
\caption{Proportion of Replications Where $X_j$ is in the Top $d$ Causative Covariates}
\begin{tabular}{|l|ccccc|}
\hline
\multicolumn{6}{|c|}{CAT-SIS} \\ \hline
 & $X_{1}$ & $X_{2}$ & $X_{3}$ & $X_{4}$ & $X_{5}$  \\ \hline\hline
$d = 10$ & 1.000 & 0.834 & 0.804 & 0.810 & 0.816 \\
$d = 15$ & 1.000 & 0.860 & 0.858 & 0.842 & 0.862 \\
$d = 20$ & 1.000 & 0.878 & 0.886 & 0.874 & 0.894 \\ \hline
\end{tabular}
\end{center}
\end{table}

\begin{table}
\begin{center}
\caption{Proportion of Replications Where $X_j$ is in the Top $d$ Causative Covariates}
\begin{tabular}{|l|ccccc|}
\hline
\multicolumn{6}{|c|}{MMLE} \\ \hline
 & $X_{1}$ & $X_{2}$ & $X_{3}$ & $X_{4}$ & $X_{5}$  \\ \hline\hline
$d = 10$ & 1.000 & 0.822 & 0.798 & 0.808 & 0.824 \\
$d = 15$ & 1.000 & 0.856 & 0.868 & 0.842 & 0.870 \\
$d = 20$ & 1.000 & 0.882 & 0.892 & 0.872 & 0.890 \\ \hline
\end{tabular}
\end{center}
\end{table}

\begin{table}
\begin{center}
\caption{Proportion of Replications Where $X_j$ is in the Top $d$ Causative Covariates}
\begin{tabular}{|l|ccccc|}
\hline
\multicolumn{6}{|c|}{DC-SIS} \\ \hline
 & $X_{1}$ & $X_{2}$ & $X_{3}$ & $X_{4}$ & $X_{5}$  \\ \hline\hline
$d = 10$ & 1.000 & 0.832 & 0.808 & 0.806 & 0.814 \\
$d = 15$ & 1.000 & 0.860 & 0.850 & 0.838 & 0.866 \\
$d = 20$ & 1.000 & 0.880 & 0.880 & 0.864 & 0.890 \\ \hline
\end{tabular}
\end{center}
\end{table}

\begin{table}
\begin{center}
\caption{Proportion of Replications Where $X_j$ is in the Top $d$ Causative Covariates}
\begin{tabular}{|l|ccccc|}
\hline
\multicolumn{6}{|c|}{HLW-SIS} \\ \hline
 & $X_{1}$ & $X_{2}$ & $X_{3}$ & $X_{4}$ & $X_{5}$  \\ \hline\hline
$d = 10$ & 1.000 & 0.742 & 0.708 & 0.710 & 0.740 \\
$d = 15$ & 1.000 & 0.794 & 0.758 & 0.778 & 0.790 \\
$d = 20$ & 1.000 & 0.832 & 0.806 & 0.800 & 0.828 \\ \hline
\end{tabular}
\end{center}
\end{table}

\subsection{Simulation 4 Results}
The results of Simulation 4 are summarized in Tables 21 through 23:
\begin{table}
\centering
\caption{Mean Minimum Model Sizes $(n = 200,~ p = 1000)$}
\begin{tabular}{ |c |  c c | }
\hline
 & CAT-SIS  & DC-SIS  \\ \hline\hline
Mean Minimum Model Size & 95.610 & 142.084 \\ 
  \hline  
\end{tabular}
\end{table}

\begin{table}
\begin{center}
\caption{Proportion of Replications Where $X_j$ is in the Top $d$ Causative Covariates}
\begin{tabular}{|l|cccccccccc|}
\hline
\multicolumn{11}{|c|}{CAT-SIS} \\ \hline
 & $X_{1}$ & $X_{2}$ & $X_{3}$ & $X_{4}$ & $X_{5}$ & $X_{6}$ & $X_{7}$ & $X_{8}$ & $X_{9}$ & $X_{10}$ \\ \hline\hline
$d = 10$ & 0.874 & 0.498 & 0.804 & 0.746 & 0.998 & 1.000 & 0.928 & 0.702 & 0.580 & 0.534 \\
$d = 15$ & 0.934 & 0.622 & 0.874 & 0.840 & 1.000 & 1.000 & 0.964 & 0.806 & 0.710 & 0.660 \\
$d = 20$ & 0.954 & 0.678 & 0.910 & 0.894 & 1.000 & 1.000 & 0.978 & 0.868 & 0.762 & 0.728 \\ \hline
\end{tabular}
\end{center}
\end{table}

\begin{table}
\begin{center}
\caption{Proportion of Replications Where $X_j$ is in the Top $d$ Causative Covariates}
\begin{tabular}{|l|cccccccccc|}
\hline
\multicolumn{11}{|c|}{DC-SIS} \\ \hline
 & $X_{1}$ & $X_{2}$ & $X_{3}$ & $X_{4}$ & $X_{5}$ & $X_{6}$ & $X_{7}$ & $X_{8}$ & $X_{9}$ & $X_{10}$ \\ \hline\hline
$d = 10$ & 0.832 & 0.444 & 0.762 & 0.678 & 0.992 & 1.000 & 0.914 & 0.648 & 0.534 & 0.468 \\
$d = 15$ & 0.900 & 0.550 & 0.838 & 0.768 & 0.998 & 1.000 & 0.950 & 0.742 & 0.640 & 0.564 \\
$d = 20$ & 0.926 & 0.620 & 0.874 & 0.828 & 1.000 & 1.000 & 0.964 & 0.796 & 0.712 & 0.650 \\ \hline
\end{tabular}
\end{center}
\end{table}

\clearpage

\section{Proofs of Theoretical Results}\label{proofSection}

Here we present in full the proofs for Theorems 1 and 2 given at \ref{Thm1} and \ref{Thm2}. Before proceeding into the proofs, we will establish a pair of lemmas which employ the Mann-Wald Theorem [see \cite{MannWald1943}]. 

\subsection{Prefacing Lemmas} These lemmas will lead into our proof of our main theorems on (strong) sure screening. 

\subsubsection{A lemma}\label{Lemma} 
\textit{[See \cite{Serfling1980}, Theorem of Section 1.7].} 

Let $\hat{\sigma}_j$ and $\hat{\sigma}_Y$ be the estimators of $\sigma_j$ and $\sigma_Y$ used in the definition of $\rhoHat$. 
Assume that $\hat{\sigma}_j$, $\hat{\sigma}_Y$, and $\hat{\tau}_j$ are all (individually speaking) \textit{consistent} estimators of the respective values they are estimating (viz. $\sigma_j$, $\sigma_Y$, and $\text{cov}(X_j, Y))$. 
Then we have that in fact \[\rhoHat = \frac{\hat{\tau}_j}{\hat{\sigma}_j\hat{\sigma}_Y}\] is a consistent estimator of $\varrho_j$. 

\begin{proof}
We will employ the Mann-Wald theorem (also known as the Continuous Mapping Theorem) twice. This theorem asserts that Borel functions that are almost everywhere continuous on $\mathbb{R}^k$ (or a Borel subset of such) preserve convergence in probability. This implies that if $\alpha$ is a consistent estimator of $A$ and $\zeta$ is a consistent estimator of $Z$, then for any Borel function $f$ satisfying the aforementioned conditions, $$f(\alpha, \zeta) \xrightarrow{p} f(A,Z)$$
and thus $f(\alpha, \zeta)$ is a consistent estimator of $f(A,Z)$.

Define the function $$f(a,b) = \frac{1}{ab}$$ on $\mathbb{R}^k_{>0} = (0,\infty)^k$ (All positive real-valued $k$-vectors). This function is continuous on its entire domain. 
(Note that, in line with condition (C1), we can assume (WLOG) that $\hat\sigma_j$ and $\hat\sigma_Y$ are both positive). 
Hence $f$ is a well defined and continuous function for operands $a = \hat\sigma_j$ and $b = \hat\sigma_Y$.
This implies by the Mann-Wald theorem that  in fact $\frac{1}{\hat\sigma_j\hat\sigma_Y}$ is a consistent estimator for $\frac{1}{\sigma_j\sigma_Y}$.  

It is taken as a given that standard binary multiplication is a Borel function on $\mathbb{R}^k$ (since multiplication is in fact continuous on all of $\mathbb{R}^k$). We implicitly use this fact above to assume that $$(\hat{\sigma}_j\hat{\sigma}_Y) \xrightarrow{p} (\sigma_j\sigma_Y).$$ Furthermore, this assumption on standard multiplication implies, again by the Mann-Wald theorem, that in fact  
\[\rhoHat = \hat{\tau}_j \frac{1}{\hat{\sigma}_j\hat{\sigma}_Y} = \frac{\hat{\tau}_j}{\hat{\sigma}_j\hat{\sigma}_Y}\] is a consistent estimator of $\varrho_j$. (Note that this result is contingent upon knowing that $\hat{\tau}_j$ is a consistent estimator of $\text{cov}(X_j, Y)$. This is to be shown below). The desired result has been achieved. 
\end{proof}


\subsubsection{Lemma on Consistency of an Estimator of Standard Deviation}\label{Lemma2}
It is a classical result (reproduced in its entirety below) that for any realizations $W_1,~ W_2, \ldots,~ W_n$ of a bounded random variable $W$,
\[S^2 = \frac{1}{n}\sum_{i = 1}^n (W_i - \bar{W})^2 \]
is a consistent estimator of $\text{Var}(W)$, where $\bar{W} = \frac{1}{n}\sum W_i$. 
As a simple corollary to this, we can once again use the Mann-Wald Theorem to get that $S$ is a consistent estimator of the standard deviation of $W$. 
\begin{proof}
Write the variance of $W$ as $\sigma^2$. It is a rudimentary result that \[\mathbb{E}(S^2) = \frac{(n-1)}{n}\sigma^2 < \sigma^2.\]
This means that $S^2$ is in fact a \textit{biased} estimator of $\sigma^2$. Let $\hat{\sigma}^2$ denote the traditional (and unbiased) estimator of $\sigma^2$:
\[\hat{\sigma}^2 = \frac{1}{n-1}\sum_{i = 1}^n (W_i - \bar{W})^2\]

It is clear that $S^2 = \frac{n-1}{n}\hat{\sigma}^2.$ It follows that
\[\text{Var}(S^2) = \text{Var}\left(\frac{n-1}{n}\hat{\sigma}^2\right) = \left(\frac{n-1}{n}\right)^2\text{Var}(\hat{\sigma}^2).\]

Furthermore, it can be established [see e.g. \cite{ChoCho2008}] that 
\[\text{Var}(\hat{\sigma}^2) = \frac{1}{n}\left(\mu_4 - \frac{n-3}{n-1}\mu^2_2\right),\]
where $\mu_\ell = \frac{1}{n}\sum(W_i - \mathbb{E}W)^\ell$ (with $\ell = 2$ or $\ell = 4$). 
Since $W$ is taken as being bounded, we know that $|\mu_\ell|<\infty$.

Employing Chebychev's inequality for any $\varepsilon>0$, we get the following:
\[\left(\mathbb{P}(|S^2 - \sigma^2|\geq \varepsilon)\right) \sim \mathbb{P}(|\hat{\sigma}^2 - \sigma^2|\geq \varepsilon) \leq \frac{\text{Var}(\hat{\sigma}^2)}{\varepsilon^2}.\]

Ergo, if we can show that the variance of $\hat{\sigma}^2$ approaches 0 as $n$ goes to $\infty$, it will follow that $S^2$ converges to $\sigma^2$ in probability (and hence is a consistent estimator of $\sigma^2$). However, we established above that \[\text{Var}(\hat{\sigma}^2) = \frac{1}{n}\left(\mu_4 - \frac{n-3}{n-1}\mu^2_2\right),\] which clearly approaches 0 as $n$ goes to infinity. This confirms that in fact $S^2$ is a consistent estimator of $\sigma^2$.
Note in conclusion that this implies by the Mann-Wald theorem that $S$ is a consistent estimator of $\sigma$.
\end{proof}

The lemma at \ref{Lemma2} establishes that indeed $\hat{\sigma}_j$ and $\hat{\sigma}_Y$ are consistent estimators of the respective standard deviations of $X_j$ and $Y$. 

We now proceed into the proofs of our main theorems on sure screening.


\subsection{Proofs of Theorems 1 and 2}

The proof of these two theorems is accomplished in three steps:
\begin{enumerate}
    \item We show that a positive lower bound $\varrho_{\min}$ exists for all $\varrho_j$ with $j \in \mathcal{S}_T$. That is, we show the following: \[\text{There exists}~~ \varrho_{\min} >0 \text{ such that } \varrho_j > \varrho_{\min}~~ \text{for all} ~~j \in \mathcal{S}_T.\]
    
    \item We then show that $\rhoHat$ is a uniformly consistent estimator of $\varrho_j$ for each $1 \leq j \leq p$. This will actually consist of showing that $\hat{\tau}_j$ is a consistent estimator of $\text{cov}(X_j, Y)$, since the terms in the denominator of $\rhoHat$ are already well established consistent estimators of the standard deviations of $X_j$ and $Y$. (Refer to the lemma at \ref{Lemma2}). 
    
    \item We finally show that there exists said constant $c > 0$ such that
    \[\mathbb{P}(\widehat{\mathcal{S}} = \mathcal{S}_T) \longrightarrow 1 \text{ as } n \longrightarrow \infty\]
    (with weak consistency being shown as a natural subcase). 
\end{enumerate}

\subsubsection{Step 1}
We know that 
\begin{eqnarray*}
    \omega_{km}^{(j)} = \left| (v_k^{(j)}-\mathbb{E}(X_j))(m-\mathbb{E}(Y)){p}_{km}^{(j)} \right|
\end{eqnarray*}                      

Hence, for $j \in \mathcal{S}_T,$
\begin{eqnarray*}
\varrho_j = \frac{\sum\limits_{\substack{1\leq k \leq K_j \\0 \leq m \leq 1}}\omega_{km}^{(j)}}{\sigma_j \sigma_Y} &\geq& \frac{1}{\sigma_{\max}^2}\sum\limits_{\substack{1\leq k \leq K_j \\0 \leq m \leq 1}}\omega_{km}^{(j)} \quad \text{by (C1),}\\
&\geq& \frac{1}{\sigma_{\max}^2} \max\limits_{\substack{1\leq k \leq K_j \\0 \leq m \leq 1}}\omega_{km}^{(j)}\\
&\geq& \frac{\omega_{\min}}{\sigma_{\max}^2} \quad \text{by (C2),}\\
&>& 0.
\end{eqnarray*}
Define $\varrho_{\min} = \dfrac{\omega_{\min}}{2\sigma_{\max}^2}.$
Then $\varrho_j > \varrho_{\min} > 0$ for all $j \in \mathcal{S_T}$.
\vspace{0.3cm}
This establishes a positive lower bound on $\varrho_j$ for all  $j \in \mathcal{S_T}$, completing Step 1. Corollary 1 at \ref{Cor1} is also established by this step.

\subsubsection{Step 2}\label{Step2}
We now need to discuss two equal forms of the numerator $\hat{\tau}_j$ of $\rhoHat$. It has been established that we desire to use $\hat{\tau}_j$ as an estimator of $\text{cov}(X_j, Y).$ We show that in fact $\hat{\tau}_j$ is equal to the following estimator for $\text{cov}(X_j, Y):$
\begin{equation}
    \frac{1}{n}\sum_{i = 1}^n(X_{ij}-\bar{X}_j)(Y_i - \bar{Y}) \approx \text{cov}(X_j, Y),
\end{equation}
where $\bar{X}_j = \frac{1}{n}\sum X_{ij}$ and $\bar{Y} = \frac{1}{n}\sum Y_i$ as before.

Specific to our case currently, we know that $Y_i \in \{0,1\}$. Assume WLOG that $X_{ij} \in \{v_{1}^{(j)},v_{2}^{(j)}, \ldots, v_{K_j}^{(j)} \}$. Then $\bar{X}_j = \bar{v}^{(j)}$. Let $n_{km}$ denote the number of observations satisfying $X_{ij} = k$ and $Y_i = m$. It follows that $\hat{p}_{km}^{(j)} = \frac{n_{km}}{n}$. We can rewrite (1) as follows:

\begin{eqnarray*}
   (1) &=& \frac{1}{n} \sum\limits_{\substack{1\leq i \leq n\\Y_i = 1}} (X_{ij}-\bar{X}_j)(1-\bar{Y})-\frac{1}{n} \sum\limits_{\substack{1\leq i \leq n\\Y_i = 0}} (X_{ij}-\bar{X}_j)(\bar{Y})\\\\
   &=& \frac{1}{n} \left((v_{1}^{(j)}- \bar{v}^{(j)})(1-\bar{Y})n_{11}  + \cdots  + (v_{K_j}^{(j)}- \bar{v}^{(j)})(1-\bar{Y})n_{K_j1} \right)\\
   && - \frac{1}{n} \left((v_{1}^{(j)}- \bar{v}^{(j)})(\bar{Y})n_{10}  + \cdots  + (v_{K_j}^{(j)}- \bar{v}^{(j)})(\bar{Y})n_{K_j0} \right)\\\\
   &=& \frac{1}{n}\sum\limits_{\substack{1\leq k \leq K_j \\0 \leq m \leq 1}}(v_k^{(j)}-\bar{v}^{(j)})(m-\bar{Y}){n}_{km}\\\\
   &=& \sum\limits_{\substack{1\leq k \leq K_j \\0 \leq m \leq 1}}(v_k^{(j)}-\bar{v}^{(j)})(m-\bar{Y})\hat{p}_{km}^{(j)}\\\\
   &=& \hat{\tau}_j.
\end{eqnarray*}
 So indeed (1) is equal to our previously given formula for $\hat{\tau}_j$. As convenient, we will use the form (1) when discussing $\hat{\tau}_j$. 
 
 We now apply the weak law of large numbers to show that $\rhoHat$ is a (uniformly) consistent estimator of $\varrho_j$. 
 This will consist of showing that $\hat{\tau}_j$ is a consistent estimator of $\text{cov}(X_j, Y)$, since the denominator of $\rhoHat$ is comprised of the routine (and, importantly here, consistent) estimators of $\sigma_j$ and $\sigma_Y$.  
 Since it can be show via a standard argument using the Mann-Wald Theorem that the quotient of consistent estimators is itself a consistent estimator, our aforementioned work with $\hat{\tau}_j$ will suffice.

By expanding the product of binomials in (1), we get 
\[\hat{\tau}_j = \frac{1}{n}\sum X_{ij}Y_i - \frac{1}{n}\sum \bar{X}_{j}Y_i - \frac{1}{n}\sum X_{ij}\bar{Y} + \underbrace{\frac{1}{n}\sum \bar{X}_j\bar{Y}}_{\rightarrow \mathbb{E}(X_j)\mathbb{E}(Y)}.\]

By several applications (summand wise) of the weak law of large numbers to this above expression, we know:
\[\frac{1}{n}\sum X_{ij}Y_i \xrightarrow{p} \mathbb{E}(X_jY)\]
\[\frac{1}{n}\sum \bar{X}_{j}Y_i \xrightarrow{p} \mathbb{E}(X_j)\mathbb{E}(Y)\]
\[\frac{1}{n}\sum {X}_{ij}\bar{Y} \xrightarrow{p} \mathbb{E}(X_j)\mathbb{E}(Y),\]
with all convergence being in probability.

Hence we have
\begin{eqnarray*}
\hat{\tau}_j &=& \frac{1}{n}\sum X_{ij}Y_i - \frac{1}{n}\sum \bar{X}_{j}Y_i - \frac{1}{n}\sum X_{ij}\bar{Y} + \frac{1}{n}\sum \bar{X}_j\bar{Y}\\
&\xrightarrow{p}& \mathbb{E}(X_jY) - 2\mathbb{E}(X_j)\mathbb{E}(Y) + \mathbb{E}(X_j)\mathbb{E}(Y)\\
&=& \mathbb{E}(X_jY) -\mathbb{E}(X_j)\mathbb{E}(Y)\\
&=& \text{cov}(X_j, Y).
\end{eqnarray*}

So indeed $\hat{\tau}_j$ is a consistent estimator of $\text{cov}(X_j, Y)$. This in turn shows, by the lemma at \ref{Lemma}, that $\rhoHat$ is consistent as an estimator of $\varrho_j$. 

It is a simple step to show that such consistency is uniform. This is done as follows:
Since $\rhoHat$ is consistent as an estimator of $\varrho_j$, we know that for any $1 \leq j \leq p$ and any  $\varepsilon > 0$, 
\[\mathbb{P}(|\rhoHat - \varrho_j| > \varepsilon) \rightarrow 0 \quad \text{ as } n\rightarrow\infty.\]

Let $J = \text{argmax}_{1 \leq j \leq p}~|\rhoHat - \varrho_j|.$
Then, since $J \in \{1,2,\ldots, p\}$ itself, we indeed know that  
\[\mathbb{P}(|\hat\varrho_J - \varrho_J| > \varepsilon) \rightarrow 0 \quad \text{ as } n\rightarrow\infty\] for any $\varepsilon> 0$. In other words, we have that 
\[\mathbb{P}\left(\max_{1 \leq j \leq p}|\rhoHat - \varrho_j| > \varepsilon\right) \rightarrow 0 \quad \text{ as } n\rightarrow\infty\] for any $\varepsilon> 0$.
This shows that $\rhoHat$ is a \textit{uniformly} consistent estimator of $\varrho_j$, completing Step 2. We also have established the claims of Corollary 2 found at \ref{Cor2}. 

\subsubsection{Step 3}
(This follows \cite{Huang2014} closely).

In Step 1 we defined $$\varrho_{\min} = \dfrac{\omega_{\min}}{2\sigma_{\max}^2}.$$ Let $c = (2/3)\varrho_{\min}$. Suppose by way of contradiction that this $c$ is insufficient to be able to claim $ \widehat{\mathcal{S}} \supseteq \mathcal{S}_T$. This would mean that there exists some $j^* \in \mathcal{S}_T$, yet $j^* \notin \widehat{\mathcal{S}}$. It then follows that we must have \[\hat{\varrho}_{j^*} \leq (2/3)\varrho_{\min}\] while at the same time having (as shown in Step 1) \[\varrho_{j^*}> \varrho_{\min}.\]

From this we can conclude that $|\hat{\varrho}_{j^*} - \varrho_{j^*}| > (1/3)\varrho_{\min},$ which implies that $\max_{1 \leq j \leq p}~|\hat{\varrho}_{j} - \varrho_{j}| > (1/3)\varrho_{\min}$ as well. 

However, we know by the uniform consistency of $\rhoHat$ that by letting $\varepsilon = 1/3 \varrho_{\min}$, we have 
\[\mathbb{P}(\widehat{\mathcal{S}} \not\supseteq \mathcal{S}_T) \leq \mathbb{P}\left(\max_{1 \leq j \leq p}|\rhoHat - \varrho_j| >(1/3)\varrho_{\min}\right) \rightarrow 0 \quad \text{ as } n\rightarrow\infty.\] This is a contradiction to the assumption of non containment above. So indeed, we have that \[\mathbb{P}(\widehat{\mathcal{S}} \supseteq \mathcal{S}_T)\rightarrow 1\quad \text{ as } n\rightarrow\infty.\]
This proves Theorem 2, and is the forward direction for proving Theorem 1.

To prove the reverse direction for Theorem 1, suppose (again by way of contradiction) that $\widehat{\mathcal{S}} \not\subseteq \mathcal{S}_T$. Then there is some $j^*\in \widehat{\mathcal{S}}$, yet $j^* \notin \mathcal{S}_T.$ This means that \[\hat{\varrho}_{j^*} \geq (2/3)\varrho_{\min},\] while at the same time (by (C2)) having \[\varrho_{j^*} = 0.\] It now follows that \[|\hat{\varrho}_{j^*} - \varrho_{j^*}| > (2/3)\varrho_{\min}.\] Set $\varepsilon = (2/3)\varrho_{\min}$. By uniform consistency we have
\[\mathbb{P}(\mathcal{S}_T \not\supseteq \widehat{\mathcal{S}}) \leq \mathbb{P}\left(\max_{1 \leq j \leq p}|\rhoHat - \varrho_j| >(2/3)\varrho_{\min}\right) \rightarrow 0 \quad \text{ as } n\rightarrow\infty.\]
From this we know that
\[\mathbb{P}(\mathcal{S}_T \supseteq \widehat{\mathcal{S}} )\rightarrow 1\quad \text{ as } n\rightarrow\infty.\]

In all, we can conclude that for $c = (2/3)\varrho_{\min}$, we have $\mathbb{P}(\mathcal{S}_T = \widehat{\mathcal{S}}) \rightarrow 1$ as $ n\rightarrow \infty$, completing the proof.

\bibliographystyle{plain}

\bibliography{BibliographyScreening}

\end{document}